# Toward a Science of Autonomy for Physical Systems: Paths

Pieter Abbeel
pabbeel@cs.berkeley.edu
University of California, Berkeley

Ken Goldberg
goldberg@berkeley.edu
University of California, Berkeley

Gregory Hager
hager@cs.jhu.edu
Johns Hopkins University

Julie Shah
julie_a_shah@csail.mit.edu
Massachusetts Institute of Technology

## 1 Introduction

An Autonomous Physical System (APS) will be expected to reliably and independently evaluate, execute, and achieve goals while respecting surrounding rules, laws, or conventions. In doing so, an APS must rely on a broad spectrum of dynamic, complex, and often imprecise information about its surroundings, the task it is to perform, and its own sensors and actuators. For example, cleaning in a home or commercial setting requires the ability to perceive, grasp, and manipulate many physical objects, the ability to reliably perform a variety of subtasks such as washing, folding, and stacking, and knowledge about local conventions such as how objects are classified and where they should be stored. The information required for reliable autonomous operation may come from external sources and from the robot's own sensor observations or in the form of direct instruction by a trainer.

Similar considerations apply across many domains – construction, manufacturing, in-home assistance, and healthcare. For example, surgeons spend many years learning about physiology and anatomy before they touch a patient. They then perform roughly 1000 surgeries under the tutelage of an expert surgeon, and they practice basic maneuvers such as suture tying thousands of times outside the operating room. All of these elements come together to achieve expertise at this task. Endowing a system with robust autonomy by traditional programming methods has thus far had limited success. Several promising new paths to acquiring and processing such data are emerging. This white paper outlines three promising research directions for enabling an APS to learn the physical and information skills necessary to perform tasks with independence and flexibility: Deep Reinforcement Learning, Human-Robot Interaction, and Cloud Robotics.

---

# 2 Deep Reinforcement Learning

As more and more data is being produced, and more and more computational power continues to become available, the opportunity to harness data towards autonomy continues to grow. In recent years computer vision and speech recognition have not only made significant leaps forward, but also rapidly increased their rate of progress, largely thanks to developments in deep learning.[19] While deep learning isn't necessarily the only way to harness the ever-growing amounts of data, it has demonstrated the ability to continue to improve performance on real world problems as more data and more compute cycles are being made available — in contrast to more traditional approaches, which have tended to saturate at some level of performance. The amount of data these deep neural nets are trained with is very large. For example, the landmark Krizhevsky et al. 2012 paper [18], which was the first to demonstrate deep learning outperform (and significantly so) more traditional approaches to computer vision on the widely referenced ImageNet benchmark, processed 200 billion images during training (obtained by shifting and re-coloring from an original labeled data-set of 1.2 million images).

Thus far the impact of deep learning has largely been in so-called supervised learning, of which image recognition and speech recognition are examples. In supervised learning one receives example inputs (e.g., images) and corresponding labels (e.g., 'cat', 'dog', etc. depending on what's in the image). The system is supposed to learn to then make correct label predictions on future inputs.

Supervised learning, however, isn't the right fit for most APS's. They aren't simply presented with a set of inputs for which they need to predict a label. Rather predictions are used to take actions which in turn will affect the next inputs encountered and so forth. This learning setting is called reinforcement learning. Preliminary results on deep reinforcement learning for Atari games at human level [24], and deep reinforcement learning visuomotor control policies [20] suggest high potential for deep reinforcement learning. However, to enable advances in autonomy for physical systems that are as transformative as what has happened over the past 5-10 years in deep supervised learning for computer vision and speech, major challenges will have to be addressed.

A first challenge is the amount of training data necessary for deep learning to succeed. Such data might be collected from passive sources such as YouTube videos, but largely will also have to encompass the APS's own experiences. As a single APS is unlikely to collect sufficient experience, learned deep representations will need to be able to leverage experience from external sources, other APS's and simulation, all well beyond what is currently possible. A second challenge is how APS's could efficiently and safely explore the vast perception-action spaces they are operating in. A third challenge is how deep learning for APS's can go beyond mapping from current percepts to actions, but also incorporate estimation, memory, and goal setting, critical for almost all practical tasks. A fourth major challenge is how to incorporate teaching and training (to be discussed in great detail in its own right in the next section) into deep reinforcement learning.

# 3 Human-Robot Interactions for Teaching and Training

APS's have traditionally been designed and deployed to work remotely from people. We now increasingly desire to integrate these systems into human environments – for example into factories, hospitals, disaster response deployments, military field operations, and in homes. The difficulty of incorporating a robot into existing work practices poses a significant barrier to the wider adoption of robot technology in these domains.

One key challenge is that many of these tasks involve skills that rely on implicit knowledge or convention that may be evident to people, but is difficult and time-consuming to explicitly capture and translate for an APS. For example, in aircraft carrier flight deck operations, veteran operators outperformed an autonomous system in generating efficient aircraft routes by using rules-of-thumb learned through training. These rules, or heuristics, captured important but implicit requirements of the domain that the autonomous system might in the future learn through observation and interaction. This opportunity exists in many other settings. An in-home APS for senior citizens can exponentially increase its utility by learning about the layout of their patron's home, their habits and preferences, and how they like their environment to be arranged. (This is, after all, what would be expected from an effective human in-home assistant.) A hospital APS will need to learn the conventions of the hospital they are in, how specific tasks are performed, and how they should interact with patients and staff. A cleaning APS needs to know where supplies are stored, how the patron likes the beds to be made, or when the children are napping and shouldn't be disturbed. A manufacturing APS for a small enterprise will need to know how to operate the tools and manipulate the stock or parts unique to that enterprise.

A second challenge arises when an APS acquires knowledge on its own through a data-driven learning process - it often lacks a ready mechanism for representing this information in a human- interpretable manner. As a result, we have limited opportunity to understand what an APS understands, and why it makes certain decisions. Without a common language or knowledge representation, we cannot rely on an APS to convey the necessary information to improve our interactions. This contributes to a lack of trust in the system. These challenges represent primary barriers towards wider utilization of APS's.

Today, we devote substantial resources to programming or teaching the APS to perform very limited skills. This is not a problem in and of itself - people often require substantial training to understand how context affects their actions, and to develop the skills and strategies to perform well under novel circumstances. Substantial benefits can be achieved in robot capability through new research directions that rethink how we teach and train robots. As tasks become increasingly complex, training helps a person or a team to establish a common understanding of the task expectations. The science of effective human training can possibly be translated to design APS's that learn to do useful work in human environments through the same techniques we use to train people. Training involves interactive observation, trial and error, and coaching and mentoring. What are the possible analogs for training machines? How would an APS interact with a mentor or teacher? How can it transfer or generalize information from prior experience to new experiences? How will these data and models be incorporated into the APS's underlying task and motion planning capabilities?

A related challenge we face is that an APS cannot readily make use of the well-honed procedures and processes we use to teach other people. One path to wider utilization of the APS, is to explore how the APS may learn using some of the same experiences and techniques that people use to gain proficiency in their work. For example, a teacher or mentor conveys knowledge through demonstration, explanation, coaching, and correction. Will teaching a physical system require all of the same elements? Will teaching require language? How will demonstration be used? Some of teaching is developing, for the student, a theory of causality around specific physical phenomena – e.g. you hold a suturing needle in this way in order to drive it effectively in this direction.

In the situation where a person and APS work cooperatively, it is just as important that the person is able to develop an effective mental model for how the APS will behave, as it is that the APS performs well in its role. How will the APS train to work alongside people? What insights can we translate from decades of study in human team training to develop new types of cooperative human-APS training procedures? Does the APS need models of its human counterparts to perform effectively, and if so, how can it learn these through interaction and training?

Finally, we note that mistakes and errors are the path toward knowledge. We expect students to make mistakes. We also expect them to, over time, to adjust their performance to eventually produce fewer mistakes, and more effective performance. Is a PAS allowed to make mistakes? What kind of mistakes? How are they identified, and how do they lead to improved performance?

## 4 Cloud-Based Robotics and Autonomy

The Cloud infrastructure and its extensive set of Internet-accessible resources has potential to enhance Autonomous Physical Systems by using data or code from a network to support operation, i.e., not all sensing, computation, and memory is integrated into a standalone system. Three potential benefits of the Cloud are: 1) Big Data: access to libraries of images, maps, trajectories, and descriptive data, 2) Cloud Computing: access to parallel grid computing on demand for statistical analysis, learning, and motion planning, 3) Collective Robot Learning: robots sharing trajectories, control policies, and outcomes.  The Cloud can also provide access to a) datasets, publications, models, benchmarks, and simulation tools, b) open competitions for designs and systems, and c) open-source software.  A survey of research in this area is available [4].  Below we summarize three potential benefits of the Cloud for autonomous physical systems.

Below we summarize three potential benefits of the Cloud for autonomous physical systems.

**Big Data:** The Cloud can provide APS's with access to vast resources of data that are not possible to maintain in onboard memory such as images, videos, trajectories, 3d models, maps, and updated data such as changes in sensor properties, traffic, weather, and floor conditions.

For example, grasping is a persistent challenge in robotics. Prior work has shown that cloud resources can facilitate incremental learning of grasp strategies [10, 26] by matching sensor data against 3D CAD models in an online database. Google Goggles [2], a free image recognition service for mobile devices, was incorporated in a prototype Cloud-based system for robot grasping [15]. Large datasets are needed to facilitate machine learning, as recently demonstrated in the context of computer vision. Large-scale image datasets such as ImageNet [11], PASCAL visual object classes dataset [13], and others [37, 40] have been used for object and scene recognition.

One research challenge is defining cross-platform formats for representing data. While sensor data such as images and point clouds have a small number of widely used formats, even relatively simple data such as trajectories have no common standards yet but research is ongoing [38, 39, 29]. Another challenge is working with sparse representations for efficient transmission of data, e.g., algorithms for sparse motion planning for robotic and automation systems [12, 21]. Also, large datasets collected from distributed sources are often "dirty" with erroneous, duplicated, or corrupted data [1, 42], such as 3D position data collected during robot calibration [23]. New approaches are required that are robust to dirty data.

**Cloud Computing:** Uncertainty in sensing, models, and control is a central issue for autonomous systems and can be modeled with numerical perturbations in position, orientation, shape, and control. Cloud Computing is ideal for sample-based Monte-Carlo analysis. For example, parallel Cloud Computing can be used to compute the outcomes of the cross product of many possible perturbations in object and environment pose, shape, and robot response to sensors and commands [41]. To facilitate sample-based methods, massively parallel computation on demand is now widely available from commercial sources such as Amazon, Google, Microsoft, and Cisco.

Cloud Computing has potential to speed up many computationally-intensive robotics and automation systems applications such as robot navigation by performing SLAM in the Cloud [32, 33] and next-view planning for object recognition [28]. Cloud-based sampling can be used to compute robust grasps in the presence of shape uncertainty [16, 17]. The Cloud also facilitates video and image analysis [35, 27], and mapping [25, 34]. Bekris et al. [7] propose an architecture for efficiently planning the motion of new robot manipulators designed for flexible manufacturing floors in which the computation is split between the robot and the Cloud.

It is important to acknowledge that the Cloud is prone to varying network latency and quality of service. Some applications are not time sensitive, such as decluttering a room or pre-computing grasp strategies or offline optimization of machine scheduling, but many applications have real-time demands.

**Collective Robot Learning:** As noted in the previous section, the Cloud can support robot learning by collecting data from many instances of physical trials and environments. For example robots and automation systems can share initial and desired conditions, associated control policies and trajectories, and importantly: data on the resulting performance and outcomes. Sharing data through Collective Robot Learning can also improve the capabilities of robots with limited computational resources [14]. The RoboEarth and RoboBrain databases are designed to be updated with new information from connected robots. The MyRobots project [3] from RobotShop proposes a "social network" for robots: "In the same way humans benefit from socializing, collaborating and sharing, robots can benefit from those interactions too by sharing their sensor information giving insight on their perspective of their current state" [5].

## 5 Challenges and Future Directions

Taken together, new learning methods, the data computation offered by the cloud, the possibility of direct learning from human instruction create new and synergistic opportunities that have never before existed. They both excite and challenge us to think in new ways, and to invent new approaches unimaginable only a few years ago.

These new paths to autonomy also introduce many challenges that will require broad participation from the computing research community. Here we list a few:

> **The Challenges of Connectivity**: It seems clear that future APS's will rely heavily on the Cloud. New algorithms and methods are needed to cope with time- varying network latency and Quality of Service. Faster data connections, both wired internet connections and wireless standards such as LTE [6], are reducing latency, but algorithms must be designed to degrade gracefully when the Cloud resources are very slow, noisy, or unavailable [8]. For example, "anytime" load balancing algorithms for speech recognition on smart phones send the speech signal to the Cloud for analysis and simultaneously process it internally and then use the best results available after a reasonable delay [9].
>
> **The Challenges of Real-world, Real-time Data Processing**: New algorithms are also needed that scale to the size of Big Data, which often contain dirty data that requires new approaches to clean or sample effectively [1, 42]. When the Cloud is used for parallel processing, it is vital that algorithms oversample to take into account that some remote processors may fail or experience long delays in returning results.
>
> **The Challenges of Mutual Trust**: Working in and around an APS will rapidly become commonplace– just as we trust the cruise control on our car, we'll come to trust its autonomous driving system. However, trust will quickly be lost if or when a human in inadvertently injured by an APS. Conversely, an APS will have a model for trust in a human – in their abilities and in their intentions. What if a human intentionally teaches an APS in a way that could lead to injury or harm?
>
> **The Challenges of Privacy**: The connectivity inherent in the Cloud raises a range of privacy and security concerns [31, 36]. These concerns include data generated by Cloud-connected robots and sensors, especially as they may include images or video or data from private homes or corporate trade secrets [44, 30]. Use of the Cloud also introduces the potential for an APS to be attacked remotely: a hacker could take over a robot and use it to disrupt functionality or cause damage. These concerns raise new regulatory, accountability and legal issues related to safety, control, and transparency [30, 22]. The "We Robot" conference is an annual forum for ethics and policy research [43].

**The Challenges of Sharing**: A potential direction for accelerating progress on APS research is to expand open source software libraries such as ROS with a model that might be called "Robotics and Automation as a Service" (RAaaS). If ROS is like Open Office, SAaaS would be like Google Docs, where software and data is installed and maintained on remote servers. This can facilitate rapid adoption and avoid problems with updates and maintenance, but introduces new challenges for security, consistency, testing and responsiveness.

These are but a few of the possible future research directions in the science of autonomy for physical systems. As the companion papers in this series illustrate, there are many unique opportunities and challenges that APS research in specific domains brings to the fore. However, we believe that with time and experience, we will find that there are common fundamental principles that will form the foundation across all domains for future autonomous systems.

# References

[1] For Big-Data Scientists, Janitor Work Is Key Hurdle to Insights. Available at: http://nyti.ms/1Aqif2X.
[2] Google Goggles. Available at: http://www.google.com/mobile/goggles/.
[3] MyRobots.com. Available at: http://myrobots.com.
[4] A Survey of Research on Cloud Robotics and Automation. Available at: http://goldberg.berkeley.edu/cloud-robotics
[5] What is MyRobots? http://myrobots.com/wiki/About.
[6] David Astely, Erik Dahlman, Anders Furuskar, Ylva Jading, Magnus Lindstrom, and Stefan Parkvall. LTE: The Evolution of Mobile Broadband. Comm. Mag., 47(4):44–51, 2009.
[7] Kostas Bekris, Rahul Shome, Athanasios Krontiris, and Andrew Dobson. Cloud Automation: Precomputing Roadmaps for Flexible Manipulation. IEEE Robotics & Automation Magazine: Special Issue on Emerging Advances and Applications in Automation, page Under Review, 2014.
[8] William Beksi and Nikos Papanikolopoulos. Point cloud culling for robot vision tasks under communication constraints. In International Conference on Intelligent Robots and Systems (IROS), 2014.
[9] Dmitry Berenson, Pieter Abbeel, and Ken Goldberg. A Robot Path Planning Framework that Learns from Experience. In International Conference on Robotics and Automation (ICRA), pages 3671–3678, May 2012.
[10] Matei Ciocarlie, Caroline Pantofaru, Kaijen Hsiao, Gary Bradski, Peter Brook, and Ethan Dreyfuss. A Side of Data With My Robot. IEEE Robotics & Automation Magazine, 18(2):44–57, June 2011.
[11] Jia Deng, Wei Dong, Richard Socher, Li-Jia Li, Kai Li, and Li Fei-Fei. Imagenet: A Large-Scale Hierarchical Image Database. In IEEE Conference on Computer Vision and Pattern Recognition, pages 248–255, 2009.
[12] Andrew Dobson, Athanasios Krontiris, and Kostas E Bekris. Sparse Roadmap Spanners. In Algorithmic Foundations of Robotics X, pages 279–296. 2013.
[13] Mark Everingham, Luc Van Gool, Christopher KI Williams, John Winn, and Andrew Zisserman. The PASCAL Visual Object Classes (VOC) Challenge. International Journal of Computer Vision, 88(2):303–338, 2010.
[14] Bruno Gouveia, David Portugal, Daniel Silva, and Lino Marques. Computation Sharing in Distributed Robotic Systems: a Case Study on SLAM. IEEE Transactions on Automation Science and Engineering (T-ASE): Special Issue on Cloud Robotics and Automation, 12(2):To appear, April 2015.
[15] B. Kehoe, A. Matsukawa, S. Candido, J. Kuffner, and K. Goldberg. Cloud-Based Robot Grasping with the Google Object Recognition Engine. In International Conference on Robotics and Automation (ICRA), 2013.
[16] Ben Kehoe, Dmitry Berenson, and Ken Goldberg. Toward Cloud-based Grasping with Uncer- tainty in Shape: Estimating Lower Bounds on Achieving Force Closure with Zero-slip Push Grasps. In International Conference on Robotics and Automation (ICRA), May 2012.
[17] Ben Kehoe, Deepak Warrier, Sachin Patil, and Ken Goldberg. Cloud-Based Grasp Planning for Toleranced Parts Using Parallelized Monte Carlo Sampling. IEEE Transactions on Automation Science and Engineering (T-ASE): Special Issue on Cloud Robotics and Automation, 12(2):To appear, April 2015.

[18] Alex Krizhevsky, Ilya Sutskever, and Geoffrey E Hinton. Imagenet classification with deep convolutional neural networks. In Advances in neural information processing systems, pages 1097–1105, 2012.

[19] Yann LeCun, Yoshua Bengio, and Geoffrey Hinton. Deep learning. Nature, 521(7553):436–444, 2015.

[20] Sergey Levine, Chelsea Finn, Trevor Darrell, and Pieter Abbeel. End-to-end training of deep visuomotor policies. arXiv preprint arXiv:1504.00702, 2015.

[21] Zheng Li, Liam O'Brien, He Zhang, and Rainbow Cai. On the Conceptualization of Performance Evaluation of IaaS Services. IEEE Transactions on Services Computing, X(X):1–1, 2014.

[22] Patrick Lin, Keith Abney, and George A. Bekey. Robot Ethics: The Ethical and Social Impli- cations of Robotics. The MIT Press, 2011.

[23] Jeffrey Mahler, Sanjay Krishnan, Michael Laskey, Siddarth Sen, Adithyavairavan Murali, Ben Kehoe, Sachin Patil, Jiannan Wang, Mike Franklin, Pieter Abbeel, and Ken Goldberg. Learn- ing Accurate Kinematic Control of Cable-Driven Surgical Robots Using Data Cleaning and Gaussian Process Regression. In IEEE International Conference on Automation Science and Engineering (CASE), 2014.

[24] Volodymyr Mnih, Koray Kavukcuoglu, David Silver, Andrei A Rusu, Joel Veness, Marc G Bellemare, Alex Graves, Martin Riedmiller, Andreas K Fidjeland, Georg Ostrovski, et al. Human-level control through deep reinforcement learning. Nature, 518(7540):529–533, 2015.

[25] Gajamohan Mohanarajah, Vladyslav Usenko, Mayank Singh, Markus Waibel, and Raffaello D'Andrea. Cloud-based Collaborative 3D Mapping in Real-Time with Low-Cost Robots. IEEE Transactions on Automation Science and Engineering (T-ASE): Special Issue on Cloud Robotics and Automation, 12(2):To appear, April 2015.

[26] M.A. Moussa and M.S. Kamel. An Experimental Approach to Robotic Grasping using a Connectionist Architecture and Generic Grasping Functions. IEEE Trans. on Systems, Man and Cybernetics, Part C, 28(2):239–253, May 1998.

[27] D Nister and H Stewenius. Scalable Recognition with a Vocabulary Tree. In IEEE Conference on Computer Vision and Pattern Recognition (CVPR), volume 2, pages 2161–2168, 2006.

[28] Gabriel Oliveira and Volkan Isler. View Planning For Cloud-Based Active Object Recognition.
Technical report, Department of Computer Science, University of Minnesota, 2013.

[29] Edson Prestes, Joel Luis Carbonera, Sandro Rama Fiorini, Vitor a. M. Jorge, Mara Abel, Raj Madhavan, Angela Locoro, Paulo Goncalves, Marcos E. Barreto, Maki Habib, Abdelghani Chibani, S´ebastien G´erard, Yacine Amirat, and Craig Schlenoff. Towards a core ontology for robotics and automation. Robotics and Autonomous Systems, 61(11):1193–1204, 2013.

[30] Andrew A Proia, Drew Simshaw, and Kris Hauser. Consumer Cloud Robotics and the Fair Information Practice Principles: Recognizing the Challenges and Opportunities Ahead. Min- nesota Journal of Law, Science & Technology, page to appear, 2014.

[31] Kui Ren, Cong Wang, Qian Wang, et al. Security Challenges for the Public Cloud. IEEE Internet Computing, 16(1):69–73, 2012.

[32] L Riazuelo, Javier Civera, and J Montiel. C2TAM: A Cloud Framework for Cooperative Tracking and Mapping. Robotics and Autonomous Systems, 62(4):401–413, 2013.
9

[33] L. Riazuelo, M. Tenorth, D. Di Marco, M. Salas, L. Mosenlechner, L. Kunze, M. Beetz, J. D. Tardos, L. Montano, and J. Montiel. RoboEarth Web-Enabled and Knowledge-Based Active Perception. In IROS Workshop on AI-based Robotics, 2013.

[34] Luis Riazuelo, Moritz Tenorth, Daniel Marco, Marta Salas, Dorian Galvez-Lopez, Lorenz Mosenlechner, Lars Kunze, Michael Beetz, Juan Tardos, Luis Montano, and J. Montiel. RoboEarth Semnatic Mapping: A Cloud Enabled Knowledge-Based Approach. IEEE Transactions on Automation Science and Engineering (T-ASE): Special Issue on Cloud Robotics and Automation, 12(2):To appear, April 2015.

[35] J. Salmeron-Garcia, F. Diaz-del Rio, P. Inigo-Blasco, and D. Cagigas. A Trade-off Analysis of a Cloud-based Robot Navigation Assistant using Stereo Image Processing. IEEE Transac- tions on Automation Science and Engineering (T-ASE): Special Issue on Cloud Robotics and Automation, 12(2):To appear, April 2015.

[36] Schmitt, Charles. Security and Privacy in the Era of Big Data.

[37] Noah Snavely, Steven M Seitz, and Richard Szeliski. Photo tourism: exploring photo collections in 3D. ACM transactions on graphics (TOG), 25(3):835–846, 2006.

[38] Moritz Tenorth and Michael Beetz. KnowRob: A Knowledge Processing Infrastructure for Cognition-Enabled Robots. International Journal of Robotics Research (IJRR), 32(5):566–590, 2013.

[39] Moritz Tenorth, Alexander Clifford Perzylo, Reinhard Lafrenz, and Michael Beetz. Representa- tion and Exchange of Knowledge about Actions, Objects, and Environments in the Roboearth Framework. IEEE Transactions on Automation Science and Engineering (T-ASE), 10(3):643–651, 2013.

[40] Antonio Torralba, Robert Fergus, and William T Freeman. 80 Million Tiny Images: A Large Data Set for Nonparametric Object and Scene Recognition. IEEE Transactions on Pattern Analysis and Machine Intelligence, 30(11):1958–1970, 2008.

[41] Jur van den Berg, Pieter Abbeel, and Ken Goldberg. LQG-MP: Optimized path planning for robots with motion uncertainty and imperfect state information. International Journal of Robotics Research (IJRR), 30(7):895–913, June 2011.

[42] Jiannan Wang, Sanjay Krishnan, Michael J. Franklin, Ken Goldberg, Tim Kraska, and Tova Milo. A Sample-and-Clean Framework for Fast and Accurate Query Processing on Dirty Data. In ACM SIGMOD International Conference on Management of Data, 2014.

[43] We Robot. We Robot Conference.

[44] Rolf H Weber. Internet of Things–New Security and Privacy Challenges. Computer Law & Security Review, 26(1):23–30, 2010.

*For citation use*: Abbeel P., Goldberg K., Hager G., & Shah J. (2015). *Toward a Science of Autonomy for Physical Systems: Paths Toward Autonomy*: A white paper prepared for the Computing Community Consortium committee of the Computing Research Association. http://cra.org/ccc/resources/ccc-led-whitepapers/

*This material is based upon work supported by the National Science Foundation under Grant No. (1136993). Any opinions, findings, and conclusions or recommendations expressed in this material are those of the author(s) and do not necessarily reflect the views of the National Science Foundation.*